**Beta Irradiation of a Geometrically Metastable Superconducting Strip Detector with a Magnetic Flux Penetration Read-Out.**

**V. Jeudy, D. Limagne, G. Waysand**

Groupe de Physique des Solides (UA 17 CNRS), Universités Paris 7 et 6
2, Place Jussieu 75251 Paris Cedex 05, France

**J. I. Collar**

Department of Physics and Astronomy, University of South Carolina,
Columbia SC 29208, U.S.A.

**T. A. Girard**

Centro de Fisica Nuclear, Universidade de Lisboa,
Av. Prof. Gama Pinto 2, 1699 Lisboa, Portugal

Geometrical metastability, observed in superconducting type I tin flat strips, has been previously proposed as a principle for particle detection. The energy deposition of an incoming β-particle induces the rupture of the metastability and consequently the penetration of multiquantum flux tubes into a superconducting tin strip. We present here the first absorption spectra from two beta sources, which demonstrate the linearity and energy-resolution of these detectors.

*PACS numbers: 74.30C, 29.40*

supr-con/9512002    10 Dec 1995

# 1. INTRODUCTION

When a superconducting strip is placed in an increasing perpendicular magnetic field, its transition to the normal state is delayed by an energy barrier that prevents spontaneous flux penetration into the volume. As suggested by J.R. Clem et al [1] and verified experimentally [2], the Gibbs free energy depends on the applied magnetic field value, the aspect ratio of the strip, and on the size of the multiquantum flux tube that may enter into the strip.

We have recently shown [3] that the nucleation of flux tubes could also be provoked by the energy deposition of incident radiation. We present here irradiation experiments on a superconducting Sn strip biased by a magnetic field, using $^{35}$S and $^{109}$Cd β sources.

# 2. GEOMETRICAL METASTABILITY : MAGNETIC NUCLEATION IN TYPE I SUPERCONDUCTING STRIPS

Geometrical metastability has been discussed in ref. [2], for tin strips of typical size 20 μm x 585 μm x 2 cm. When a superconducting strip is subject to a magnetic field applied perpendicular to the strip surface, $H_{app}$, its transition to the normal state crosses three different penetration regimes [4]:

- regime 1 : for $H_{app} < H_{fip}$ (first irreversible flux penetration), the flux penetration is reversible and therefore controlled by the ramping rate of the magnetic field. An intermediate state structure installs itself on the sharp edges of the superconducting strip due to the edge effect [5]. The intermediate state has a curtainlike structure whose folds are perpendicular to the strip border.

- regime 2 : for $H_{fip} < H_{app} < H_{lip}$ (last irreversible flux penetration), the penetration becomes irreversible. The transition is achieved through the penetration of multiquantum flux bundles, nucleated on the strip edges.

As discussed in ref. [1] and [2], the first penetrating flux bundle is nucleated by the pinching off of a fold in the intermediate state present on the strip edges. We have shown that this bundle has a surface proportional to the square of the fold width [6], i.e., proportional to the coherence length.

During the magnetic field increase, the nucleated flux bundles move toward the strip center [4] and create an intermediate state structure. Photographs, obtained by magneto-optical techniques [7] on flat superconducting strips, present a global inhomogeneity of the flux distribution on the surface of the sample. For long superconducting strips, the intermediate state present in the strip center does not reach the strip borders, where two perfectly diamagnetic bands reflect the presence of the energy barrier. As the magnetic field is increased, the surface occupied by the intermediate state structure expands towards the strip borders. Simultaneously, the width of the diamagnetic bands is reduced.

- regime 3 : for $H_{lip} < H_{app} < H_c$ (the thermodynamical critical field at the operating temperature), the flux penetration becomes again reversible. The two diamagnetic bands disappear and the intermediate state occupies all the strip area.

## 3. RUPTURE OF THE METASTABILITY BY THERMAL HEATING

### 3.1. Thermal heating

The rupture of metastability (regime 2) can be achieved by thermal nucleation. The energy deposition of an incident particle produces local heating that can induce the penetration of flux bundles. However, due to the inhomogeneity of the global flux structure in the strip, the thermal nucleation efficiency depends on the position of the heated zone. Since the energy deposition is only detected if it induces a flux jump in the pick-up coil, a signal may not be recorded if a particle loses its energy far from a flux bearing zone. In other words, the particle has to release heat near the strip edges, where the energy barrier is present, to induce flux penetration.

### 3.2. Experimental set up

The strip was cut from a Goodfellow 99.9% pure annealed, free of pin holes, 20 μm thick ($L_y$) tin foil. Its width and length are 760 ± 30 μm and 2.4 ± 0.1 cm, respectively. The $^{35}$S source (100% $\beta^-$ decay, $Q_{\beta^-}$ = 167.5 keV) was made by evaporating an aqueous solution onto a 20 μm thick smoking paper. The total source activity is 5.6 MBq, spread over the 1 cm x 3 cm paper surface. The $^{109}$Cd source was made by evaporating an HCl solution. Its activity is 43 kBq, spread over the 0.85 cm x 2.85 cm paper surface. One side of the superconducting strip was totally covered by the source; the other side was placed against a U-shaped detection loop.

The flux penetration is detected by the fast acquisition system employed with metastable superconducting granule detectors [8]. The nucleation of flux bundles creates discontinuities in the flux cut by the detecting loop, which is connected via two pulse transformers to a LeCroy HQV810 based pulse amplifier. This system detects irreversible flux variations only, i.e., those controlled by the damping of the eddy currents, whereas reversible flux changes are not detected. The input signal is then shaped with a fast LeCroy amplifier. For the amplitude analysis, the amplifier output is branched to the input of a 10 bit multichannel analyser (LeCroy qVt 3001).

The experiments were performed in an $^4$He cryostat equipped with a single shot $^3$He minifridge. The $^3$He bath stably cooled a copper plate at a temperature of 400 mK.

### 3.3. $^{35}$S absorption spectrum

The magnetic field, applied perpendicularly to the strip, was ramped up from zero at a rate of 16 G / s. The ramping was stopped for 10 seconds at a pause field of 65 G, where the highest number of transitions due to irradiation was detected. The magnetic field ramping was then resumed to 500 G, well above $H_c$ (390 mK), and returned to zero.

During the pause, no transitions were detected after 8 s: the detector is saturated and must be reset for new particle detection. To obtain the experimental energy absorption spectrum, the procedure was repeated 30 times to improve the statistics, and the 1024 channels were grouped by 16. The experimental curve (Fig. 1: •) exhibits a hump centred around channel # 300.

A Monte Carlo simulation was performed to obtain an estimate of the electron energy loss in the strip and to identify the origin of the hump. Fig. 1 shows a preliminary comparison between the pulse-height spectrum and the simulation of the electron energy deposition in the strip. The simulation includes energy losses at the source, Molière straggling [9] and backscattering at the source-detector interface using the model of Archard [10]. Energy loss at both source and strip are computed via the Bethe-Bloch equation. To calibrate the pulse-height spectrum, a linear fit was made to the region close to the last occupied channel (# 971) and the obtained end-point was simply adjusted to the Q-value of the β emission spectrum (167.5 keV). The excess of counts at channel # ~ 300 and of deposited energy at ~ 45 keV then coincide. This excess comes from the fact that electrons emitted with energies higher than ~ 70 keV can cross the strip thickness (20 μm) without losing all their energy. This together with backscattering promotes partial energy deposition. We found a dependence of the simulated energy deposition on the precise theoretical model of Molière straggling employed. The backscattering theory of Archard gave the best agreement between experiment and simulation.

### 3.4. $^{109}$Cd absorption spectrum

The superconducting strip was subject to a magnetic field increasing continuously, from zero, at a rate of 7 G / s, up to 500 G. The magnetic field was then decreased back to 0 G.

Fig. 2 represents the pulse-height spectrum recorded during the sweeping up of the magnetic field. The experimental curve exhibits a noise peak centred on channel # 160, followed by two other peaks, centred on channel # 450 and # 600, corresponding to the two main β emissions of the $^{109}$Cd source, 62 keV (42%) and 84 keV (44%), respectively. The counts in higher channels are not induced by energy deposition, but arise from magnetic nucleation during the sweeping up.

The equality between the ratio of the two emission energies (84 keV / 62 keV = 1.3) and the ratio of the corresponding channel numbers (600 / 450 = 1.3), together with the results obtained with the $^{35}$S source, show that the amplitude of the detected signal varies linearly with the deposited energy. Since the time variation of the magnetic flux in the coil, $d\phi / dt$, is integrated by the electronics, the recorded pulse-amplitude is proportional to the flux contained in the thermally nucleated normal tube, i.e., pulse-height ~ $\int (d\phi / dt) \, dt \sim \phi = H_c(T) \, S_{tube}$, where $S_{tube}$ is the surface area of the flux tube. The observed linearity of the spectrum then implies that the nucleated volume $L_y \, S_{tube}$ must be proportional to the energy deposited by a particle at a given temperature:

$$\frac{E_{dep}}{L_y \, S_{tube}} = \text{constant.} \tag{1}$$

Fig. 2 represents the first energy-resolved particle spectrum obtained with a geometrically-metastable detector. The energy resolution is ~19 % (FWHM) at 62 keV, comparable to that of NaI. Energy losses at the source are however largely responsible for the observed FWHM. We plan to substitute the source by an improved, much thinner one.

## 4. CONCLUSIONS

The transition of a superconducting strip subject to a continuously increasing perpendicular magnetic field is achieved by the penetration of multiquantum flux bundles nucleated on the edges of the strip.

The thermally nucleated flux bundles have a volume which is proportional to the energy absorbed under irradiation. Unlike in current-biased strip detectors [11], a particle doesn't have to heat the full strip or its flux cross section to be detected.

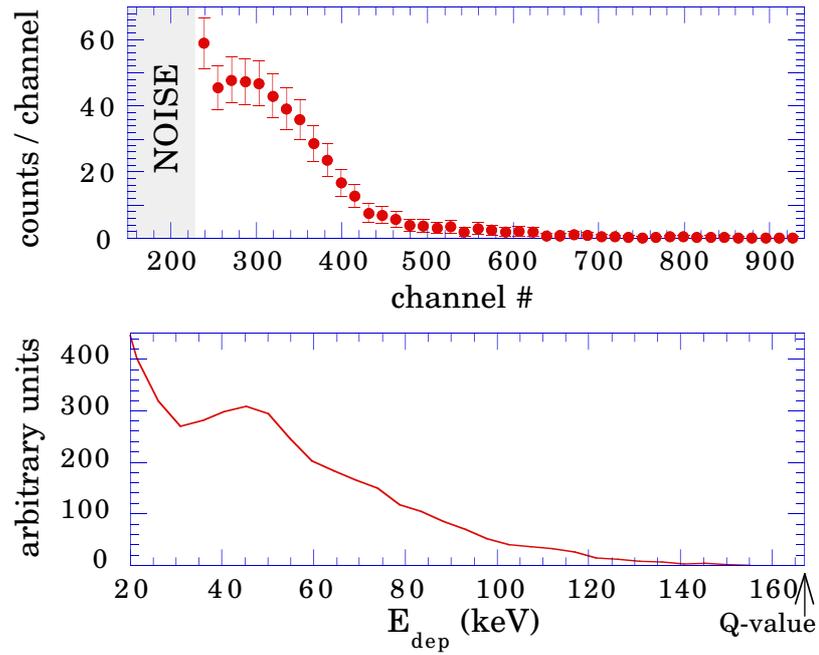

Fig. 1.

Top figure: sum, over 30 runs, of the number of counts recorded during the 10 s pauses; T= 390 mK. Bottom figure: the simulated absorption spectrum, calibrated so that the Q-value coincides with the end-point of the experimental spectrum.

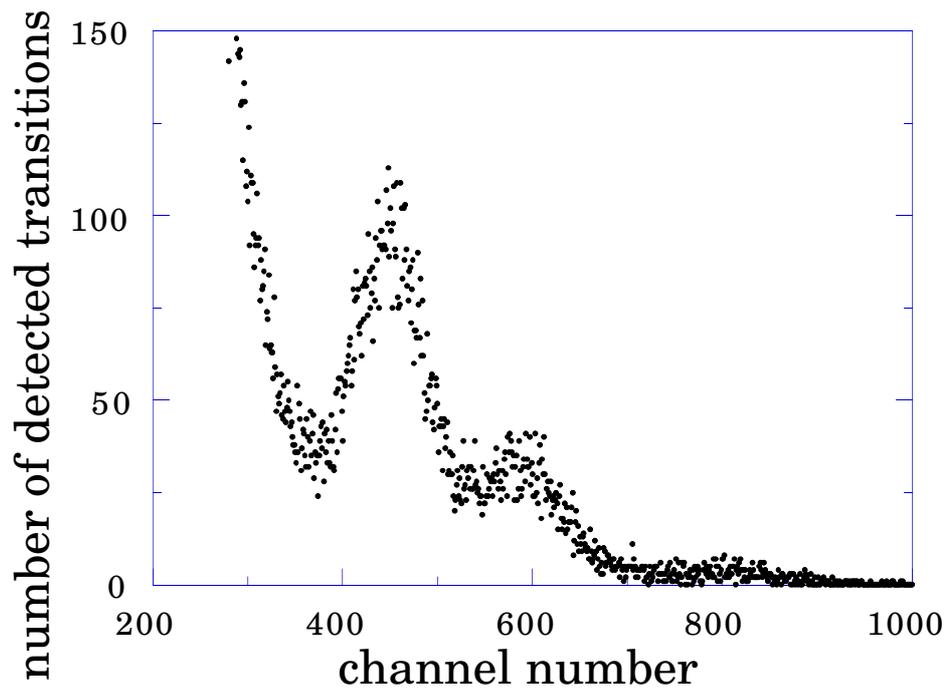

Fig. 2.
Pulse-height spectrum from a $^{109}$Cd source. The number of counts corresponds to the sum over 20 runs. T= 400 mK.